\documentclass[preprint]{emulateapj} 

\usepackage{natbib}
\usepackage{hyperref}
\hypersetup{colorlinks,citecolor=Blue,linkcolor=Red,urlcolor=Blue}
\usepackage{amsmath}
\usepackage{empheq}
\usepackage[usenames,dvipsnames]{color}
\usepackage{enumitem}
\setlist{noitemsep}

\allowdisplaybreaks 

\newcommand{\appropto}{\mathrel{\vcenter{\offinterlineskip\halign{\hfil$##$\cr\propto\cr\noalign{\kern2pt}\sim\cr\noalign{\kern-2pt}}}}}

\newcommand{\G}{\mathcal{G}}
\newcommand{\Rh}{R_{\rm{H}}}
\newcommand{\Mstar}{M_{\star}}
\newcommand{\Mp}{M}
\newcommand{\Rp}{R}
\newcommand{\Mdot}{\dot{M}}

\begin{document}
 
\title{On the Terminal Rotation Rates of Giant Planets}

\author{Konstantin Batygin$^1$} 
\affil{$^1$Division of Geological and Planetary Sciences, California Institute of Technology, 1200 E. California Blvd., Pasadena, CA 91125}

\begin{abstract}
Within the general framework of core-nucleated accretion theory of giant planet formation, the conglomeration of massive gaseous envelopes is facilitated by a transient period of rapid accumulation of nebular material. While the concurrent buildup of angular momentum is expected to leave newly formed planets spinning at near-breakup velocities, Jupiter and Saturn, as well as super-Jovian long-period extrasolar planets, are observed to rotate well below criticality. In this work, we demonstrate that the large luminosity of a young giant planet simultaneously leads to the generation of a strong planetary magnetic field, as well as thermal ionization of the circumplanetary disk. The ensuing magnetic coupling between the planetary interior and the quasi-Keplerian motion of the disk results in efficient braking of planetary rotation, with hydrodynamic circulation of gas within the Hill sphere playing the key role of expelling spin angular momentum to the circumstellar nebula. Our results place early-stage giant planet and stellar rotation within the same evolutionary framework, and motivate further exploration of magnetohydrodynamic phenomena in the context of the final stages of giant planet formation.
\end{abstract}

\section{Introduction}\label{sect1}
A defining characteristic of the solar system's planetary album is the presence of gas giants - Jupiter and Saturn. Possessing the dominant share of the solar system's angular momentum budget, it is firmly established that the formation and early evolution of this pair of objects is deeply implicated in shaping the solar system's remarkable present-day architecture \citep{Tsiganis2005,Walsh2011,BL2015}. Placing Jupiter and Saturn into their greater Galactic context, the aggregate of confirmed extrasolar planets has revealed that Jovian-class planets span a staggering extent of stellocentric radii, with orbital periods ranging from fractions of a day to hundreds of years. However, in face of their pronounced presence within the current planetary census (as well as the critical roles they play in sculpting their orbital neighborhoods), the physical processes that regulate the final stages of giant planet conglomeration remain imperfectly understood. 

Within the broadly accepted framework of core-nucleated accretion theory \citep{BodPol1986,Pollack1996}, the process of giant planet formation is envisioned to unfold as a sequence of three phases. The first phase corresponds to the formation of a high-metallicity core comprising $M_{\rm{core}}\sim10-20\,M_{\oplus}$ of icy/rocky material. During the second phase, this core acquires a gaseous hydrostatic envelope, whose slow growth is facilitated by cooling and the associated Kelvin-Helmholtz contraction. Upon reaching a mass comparable to that of the core, the envelope enters the third phase of conglomeration, characterized by runaway accretion of the nebular gas. In this narrative, two issues remain elusive: what sets (1) the final mass and (2) the terminal rotation rate of the newly-formed giant planet? In this work, we focus primarily on the latter problem.

Naively speaking, one may expect that during the final stage of core-nucleated conglomeration, a planet is bound to accrete the angular momentum of the nebular gas along with its mass, resulting in a terminal rotation rate close to breakup velocity (wherein the surface layers rotate at essentially the orbital speed). Subsequent gravitational contraction of the planet would only exacerbate this problem. In contrast with this view, the $9.93\,$hour and $10.7\,$hour spin periods of Jupiter and Saturn lie significantly below the corresponding breakup values (28\,\% and 37\,\% respectively). While the surprisingly slow rotation of Jupiter and Saturn was appreciated as a theoretical puzzle even before the wide-spread detection of extrasolar planets \citep{TS1996}, the first constraints on the rotation rates of long-period planetary mass companions presented by \citet{Bryan2017} have shown that strongly sub-breakup rotation rates are in fact the rule, rather than the exception. In particular, normalized by corresponding breakup velocities, the rotation rates of the five objects considered by \citet{Bryan2017} confidently lie below $\sim0.5$, with the $0.05-0.3$ range appearing characteristic. Moreover, the same study has demonstrated that the measured rotation rates do not exhibit a statistically significant dependence on age, suggesting that the mechanism responsible for setting terminal spin rates of giant planets operates during their infancy.

Recent theoretical analyses of the final stages of giant planet formation have shown that contrary to traditional 1D models \citep{Dave1982,Pollack1996,Bat2016}, the infall of nebular gas becomes spherically asymmetric once the planet acquires a sufficient amount of mass to open a gap in its natal protoplanetary disk \citep{Crida2006,Fung2014}. That is, high-resolution nested grid hydrodynamic simulations of \citet{Tanigawa2012,Gressel2013,Judit2016} have shown that a strong meridional circulation of gas develops within the planet's Hill sphere, wherein disk material precipitates down towards the planet from high altitudes, generating a quasi-Keplerian circumplanetary disk (which in turn feeds the gas back to the mid-plane of the circumstellar nebula). As pointed out by \citet{Judit2016}, this picture further aggravates the terminal spin problem, since angular momentum exchange between the circumplanetary disk and the planet is generally expected to increase the rotation rate of the planet.

Qualitatively, the terminal spin problem of giant planet formation is reminiscent of the low angular momentum problem of T-Tauri stars, which are also observed to rotate substantially slower than initially expected (see \citealt{Bouvier2013} for a review). To this end, it has been shown that magnetic coupling between the star and the inner regions of the circumstellar disk can broadly explain the diminished rotation rates of newborn stars \citep{Konigl1991,MattPudritz2005}. A similar process has been suggested within the context of Jovian formation \citep{TS1996,Turner2014}, and here we will revisit magnetohydrodynamic interactions between a young giant planet and its circumplanetary disk in light of recent theoretical developments. In particular, the calculations presented below demonstrate that the process of hydromagnetic planet-disk coupling naturally leads to slow planetary rotation rates, with rapid recycling of gas within the Hill sphere playing a crucial role in the envisioned mechanism. The remainder of the paper is organized as follows. In section (\ref{sect2}), we present our semi-analytical model. In section (\ref{sect3}), we compute the secular spin evolution of a long-period, Jupiter-mass planet subject to magnetohydrodynamic effects. We summarize and discuss our results in section (\ref{sect4}).

\section{Model}\label{sect2}
As a starting point, let us outline the key ingredients of our semi-analytical model. Given that direct observational constraints on the giant planet formation process remain stubbornly elusive, here we will focus primarily on characterizing the physical processes at play, rather than any specific scenario. Thus, simplicity will be emphasized wherever possible, and in contrast with numerous state-of-the-art numerical experiments available in the literature, the input parameters we employ should essentially be viewed as order-of-magnitude estimates (most relevant to a Jupiter-mass planet residing beyond the ice-line of its natal protoplanetary nebula). We begin with a description of the circumplanetary disk. 

\subsection{Circumplanetary Disk}
\begin{figure}
\includegraphics[width=\columnwidth]{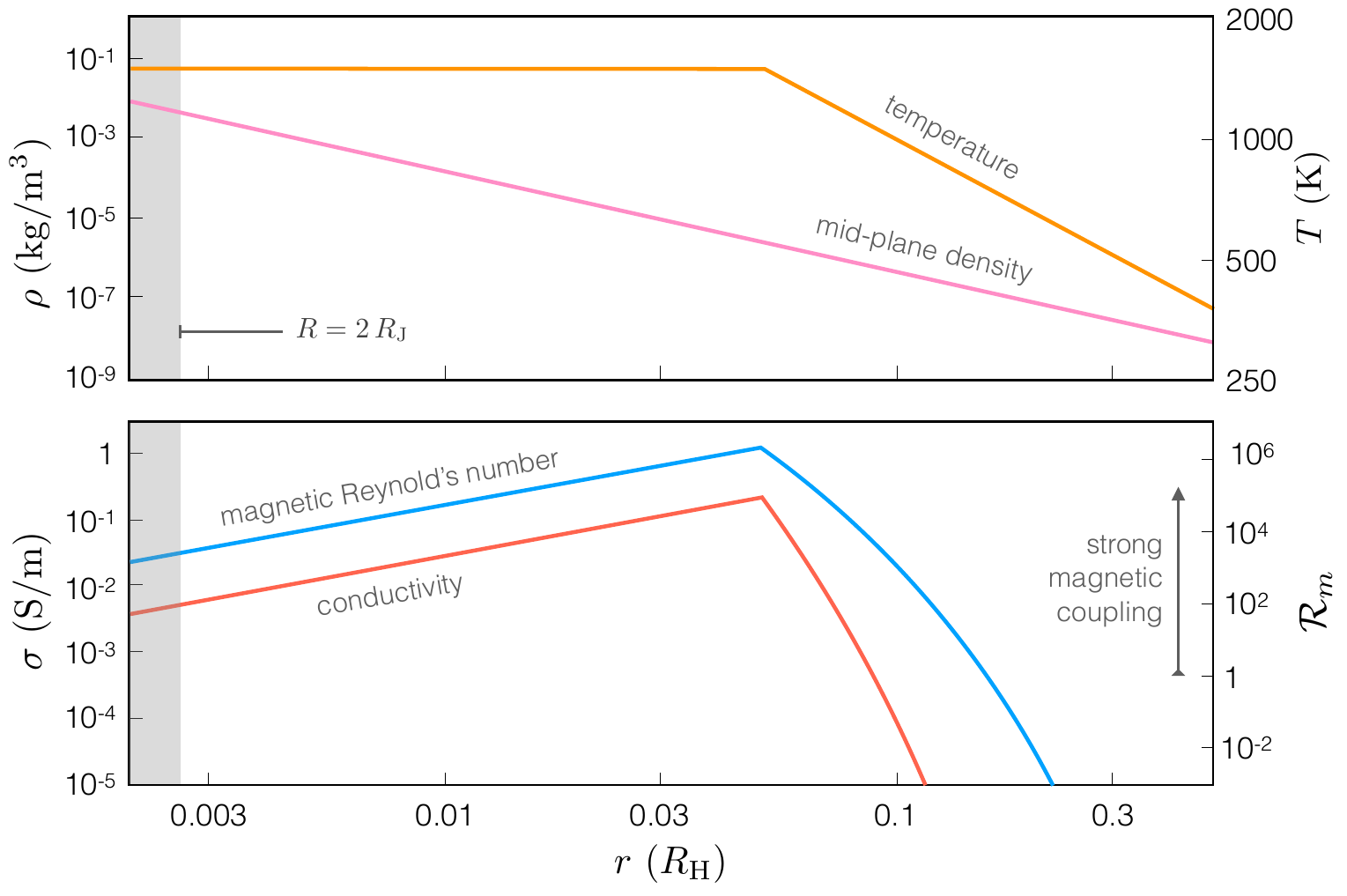}
\caption{Structure of the circumplanetary disk. The top panel shows the assumed temperature (orange) and mid-plane density (pink) profiles. Owing to a temperature profile shallower than $T\propto1/r$, the disk is strongly flared, with an aspect ratio ranging from $\sim0.1$ at its inner boundary to $\sim0.6$ at a third of the Hill radius. The bottom panel depicts the computed electrical conductivity profile within the disk (red), as well as the corresponding magnetic Reynolds number (blue). Note that interior of $r\lesssim0.15\Rh$, the system lies in the regime of strong magnetic coupling.}
\label{diskfig}
\end{figure}

A standard approach to modeling astrophysical disks is to assume that their temperature, $T$, and mid-plane density\footnote{We remind the reader that in a vertically isothermal disk, the more commonly used surface density, $\Sigma$, is related to the mid-plane density via $\rho=\Sigma/\sqrt{2\pi}\,h$, where $h$ is the disk scale-height.}, $\rho$, profiles take the form of power-laws in orbital radius \citep{Armitage2010}. Here, we follow this convention and adopt the following functional forms:
\begin{align}
T &= \min\bigg[ T_{\rm{max}}, T_0\, \bigg( \frac{\Rh}{r} \bigg)^{3/5} \bigg] \nonumber \\
\rho &= \rho_0 \, \bigg( \frac{\Rh}{r} \bigg)^{5/2},
\label{Trho}
\end{align}
where $\Rh = a\,(\Mp/3\Mstar)^{1/3}=0.36\,$AU is the Hill radius, $T_{\rm{max}}=1500\,$K, $T_0=250\,$K, and $\rho_0=1.5\times10^{-9}\,$kg/m$^3$ (Figure \ref{diskfig}). These simple relations provide a good match to the simulated disk of \citet{Judit2016,Judit2017}, where a Jupiter mass planet is held at a constant surface temperature of $1500\,$K. We note, however, that unlike the quoted hydrodynamical simulations, our broken power-law temperature profile does not rise above $T_{\rm{max}}$, yielding an isothermal disk interior to $r<0.05\,\Rh$ (as we will see below, our analysis is insensitive to the exact functional form of $T$ at sufficiently high temperatures).

Although conditions attained within a disk parameterized by equations (\ref{Trho}) are insufficient to generate Hydrogen or Helium ions in any appreciable fraction, temperatures in excess of $T\gtrsim1000\,$K are high enough to thermally ionize alkali metals, which are expected to be present in trace amounts within the nebula \citep{Armitage2010}. Accordingly, assuming that the disk material is an ideal gas of solar composition, we have computed the ionization levels of Na, K, Li, Rb, Fe, Cs, and Ca, employing the Saha equation:
\begin{align}
\frac{n_i^+\,n_e}{n_i-n_i^+}=\bigg(\frac{m_e\,k_{\rm{b}}\,T}{2\pi\,\hbar^2} \bigg)^{3/2}\exp\big(-I_i/k_{\rm{b}}\,T\big),
\label{Saha}
\end{align}
where $n_i^+$ and $n_i$ are positive ion and total number densities of constituent $i$, $n_e$ is the total electron number density, $m_e$ is the electron mass, $k_{\rm{b}}$ and $\hbar$ are Boltzmann's and Plank's constants, and $I_i$ is the ionization potential of constituent $i$. The appropriate elemental abundances and ionization potentials were adopted from \citet{Lodders1999} and \citet{Cox2000}, respectively. The electrical conductivity within the circumplanetary disk was then computed using the standard expression
\begin{align}
\sigma=\frac{n_e}{n_n}\frac{e^2}{m_e\,\mathcal{A}_c}\sqrt{\frac{\pi\,m_e}{8\,k_{\rm{b}}\,T}},
\label{cond}
\end{align}
where $e$ is the electron charge, $n_n$ is the number density of neutral particles, and $\mathcal{A}_c$ is the number density weighted scattering cross section of Hydrogen and Helium.

The electrical conductivity profile of the disk is shown in Figure (\ref{diskfig}). Within the inner $\sim5\%$ of the Hill radius, $\sigma$ ranges from $\sim0.005-0.3\,$S/m - a value 1-2 orders of magnitude smaller than the conductivity of salty water. Crucially, however, the conductivity itself is not the primary quantity of interest. Instead, magnetic induction within the disk is characterized by the magnetic Reynold's number \citep{Moffatt1978}:
\begin{align}
\mathcal{R}_m=\frac{v_{\rm{kep}}\,h}{\eta}=\mu_0\,\sigma\,r\,\sqrt{\frac{k_{\rm{b}}\,T}{\mu}},
\label{Rem}
\end{align}
where $\eta=1/\mu_0\,\sigma$ is the magnetic diffusivity, $\mu_0$ is the permeability of free space and $\mu=2.4\,m_{\rm{H}}$ is the mean molecular weight of the gas. As shown in Figure (\ref{diskfig}), magnetic Reynold's number exceeds unity by orders of magnitude out to orbital distances comprising a substantial fraction of the Hill radius. More specifically, $\mathcal{R}_m\gg1$ for $r\lesssim0.15\,\Rh$. Therefore, even if the disk is substantially sub-Keplerian (as is the case, for example, in the simulations of \citealt{Judit2016}, where $v/v_{\rm{kep}}\approx 0.8$) the system is guaranteed to reside in a regime where magnetic diffusion plays an essentially negligible role on orbital timescales.

\subsection{Planetary Dynamo}
With a rudimentary description of the circumplanetary disk in place, let us now consider the planetary (i.e. central body) parameters. As already mentioned above, we are interested in quantifying the final stages of conglomeration of a Jupiter-mass object. Correspondingly, following \citet{Lissauer2009,BCumming2017}, here we adopt a mass-accretion rate of $\Mdot=10^{-3}\,M_{\oplus}/$yr, a radius of $\Rp=2R_{\rm{J}}$, and an effective temperature of $T_{\rm{eff}}=1500\,$K, in agreement with the energy boundary condition of the simulated disk in \citet{Judit2016}. Cumulatively, these parameters yield a proto-Jovian luminosity of $L_p\approx 2\times 10^{-4}\,L_{\odot}$, and roughly correspond to the so-called ``hot start" initial conditions of giant planet evolution.

A direct consequence of convective energy transport within the planetary interior is the generation of a large-scale magnetic field. Correspondingly, \citet{Christensen2009} have proposed the following equipartition-like relationship that connects the interior heat flux with the magnitude of the generated field:
\begin{align}
\langle B^2 \rangle/2\mu_0 = c\,f_{\rm{ohm}}\, \langle \rho \rangle^{1/3}\,\big(F\,q \big)^{2/3} \sim \langle \rho \rangle\,v_{\rm{conv}}^2.
\label{Eflux}
\end{align}
In the above expression, $c$ is a constant of proportionality of order unity, $f_{\rm{ohm}}\approx1$ is the ratio of Ohmic to total energy dissipation, $\langle\rho\rangle$ is the average density of the field generating region, $q=\sigma_{\rm{sb}}\,T_{\rm{eff}}^4$ is the bolometric flux, and $F$ is a quantity related to the ratio of the largest convective length-scale to the temperature scale-height. Furthermore, the surface field is related to the volume-averaged dynamo field through a simple numerical factor $\langle B\rangle/B_{\rm{s}}\approx3.5$.

Intriguingly, the relationship (\ref{Eflux}) successfully connects the Geodynamo, Jupiter's dynamo, as well as fully convective stars, rendering our estimate of the proto-Jovian field an interpolation. For the present-day Jupiter, \citet{Christensen2009} advocate for $F=0.35$, which yields $c=1.18$. Retaining these numbers, we obtain $B_{\rm{s}}\approx500\,$Gauss as an estimate for our model planetary surface field strength - a quantity approximately two orders of magnitude in excess of Jupiter's present-day field. Although direct measurements of newly-formed giant planets' magnetic fields do not yet exist, it is worth noting that our inferred value is almost an order of magnitude smaller than the recently derived magnetic fields of $10-30\,M_{\rm{J}}$ brown dwarfs characterized by $T_{\rm{eff}}\sim1000\,$K \citep{Kao2016}, suggesting that a slightly sub-kiloGauss field is probably not an unreasonable estimate for a highly luminous giant planet (see also \citealt{Stevenson1974}).

\section{Results}\label{sect3}

\begin{figure*}
\includegraphics[width=\textwidth]{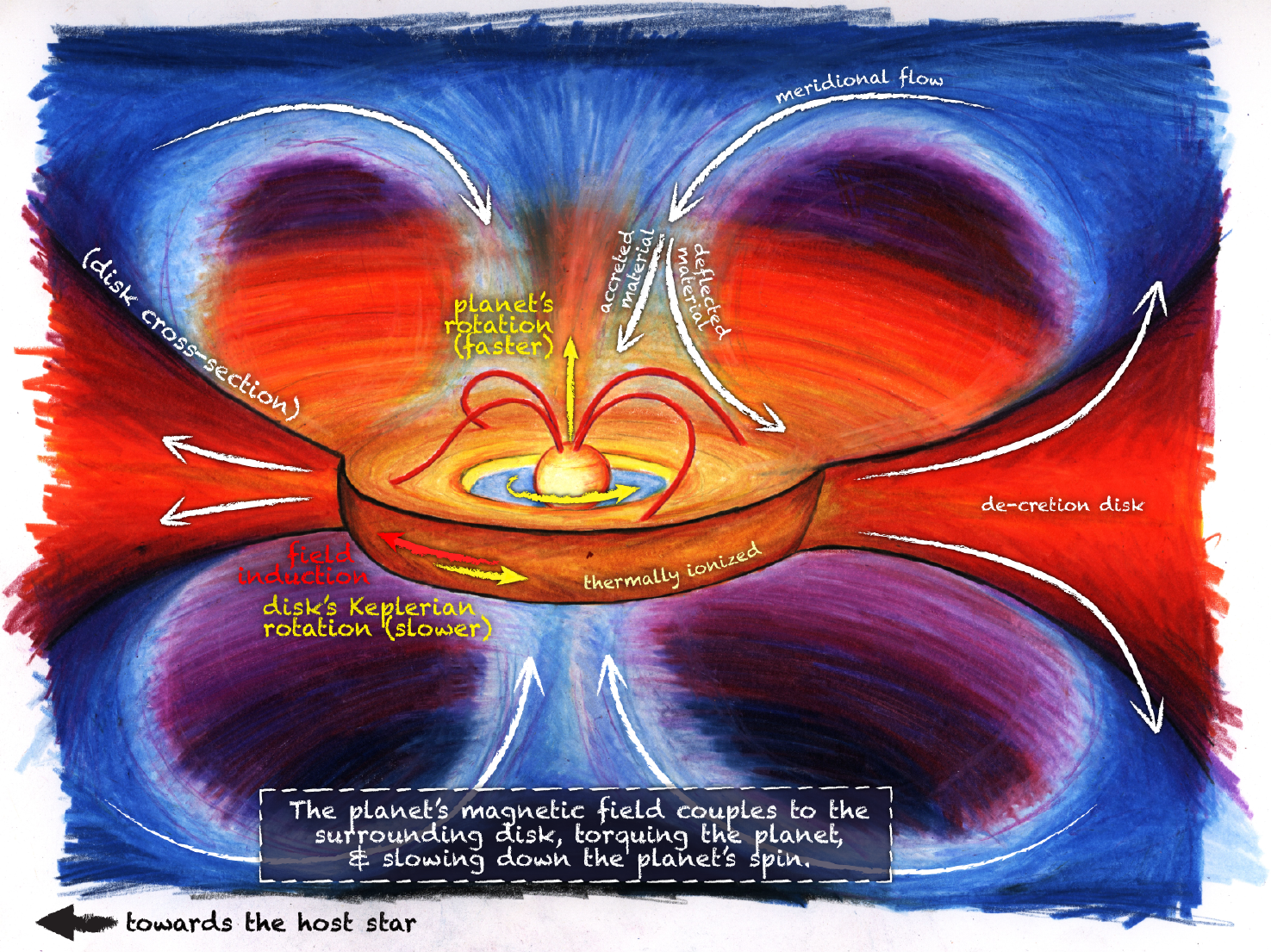}
\caption{A newly-formed giant planet embedded within the circumstellar disk. Meridional flow of gas within the Hill sphere feeds a circumplanetary de-cretion disk that connects back onto the protoplanetary nebula. Thermal ionization of Alkali metals ensues in the inner regions of the disk, coupling the planetary magnetic field to the quasi-Keplerian motion of the gas. Consequently, spin angular momentum is expelled into the circumplanetary disk, resulting in a secular decay of the planetary rotation rate.}
\label{pretty}
\end{figure*}

With the primary components of the model delineated in the previous section, we may now consider the consequences of interactions between the strongly magnetized planet and the electrically conductive circumplanetary disk. We begin by examining the effects of the planetary magnetic field upon the accretionary flow in its vicinity, and subsequently evaluate the planetary spin evolution subject to magnetohydrodynamic disk-planet coupling.

\subsection{Magnetic Truncation}
Provided an accretionary flow of magnitude $\Mdot$, a characteristic radial scale interior to which magnetic effects dominate the dynamics of the system is given by
\begin{align}
R_t\sim\bigg(\frac{\pi^2}{2\mu_0}\frac{\mathcal{M}^4}{\G\,\Mp\,\Mdot^2} \bigg)^{1/7},
\label{Rtr}
\end{align}
where $\mathcal{M}=B_{\rm{s}}\,\Rp^3$, and the field configuration is assumed to be a pole-aligned dipole\footnote{While here we opt for a simple field configuration for definitiveness, we note that the model does not require any particular geometry to operate.}. This well-known expression has been derived by various means in the literature, including balance of the magnetic and ram pressure \citep{GhoshLamb1979} as well as considerations of angular momentum transport \citep{OstrikerShu1995}. However, as pointed out by \citet{MohantyShu2008}, apart from a numerical coefficient of order unity, any physical system determined by the quantities inside the parenthesis must have a characteristic length-scale given by equation (\ref{Rtr}) by dimensional analysis. Given our fiducial parameters quoted in the previous section, we obtain $R_t\sim4-5\,R_{\rm{J}}\approx0.006\,\Rh$.

For radii interior to $R_t$, the accretionary flow is envisioned to take on a force-free configuration, wherein $\mathbf{v}\times\mathbf{B}\rightarrow0$ \citep{MattPudritz2005,MohantyShu2008}. That is, instead of free-falling, the weakly-ionized gas becomes confined to the magnetic field lines, forming an ``accretion curtain" around the newly formed planet. While this picture appears qualitatively identical to the one typically invoked for disk-bearing T-Tauri stars (see e.g. \citealt{AG2012}), it is crucial to note that the flow direction is reversed. 

In the case of young stars, the disk has an overall inward radial velocity, and upon reaching the truncation radius, the gas climbs up the critical field line, eventually arriving to the vicinity of the stellar pole \citep{BatAd2013}. In the case of the circumplanetary disk, the accretion flow is precipitating upon the planet from a high altitude, implying nearly vertical motion \citep{Tanigawa2012,Judit2016}. Thus, upon impacting the critical field line, the gas either accretes upon the planet (if it lands interior to the apex of the critical field line) or slides down the magnetic curtain to join the circumplanetary disk. In the latter case, the gas is eventually expelled from the Hill sphere as circumplanetary disk material is recycled back into the nebula \citep{Morby2014}. 

The picture described above implies that upon becoming luminous enough to develop a field that is sufficiently strong to partially isolate the vertically precipitating flow (i.e. $R_t\gtrsim\,2R_{\rm{J}}$), a young planet begins to magnetically limit the geometric cross-section through which it can accrete nebular material (Figure \ref{pretty}). Crudely speaking, for a dipole field, this entails a vertical accretion cross-section of $\mathcal{A}_{\rm{accr}}\approx \pi R_t^2/3^{3/4}$. Therefore, it is plausible that this effect may play an important role in quenching the conglomeration rate of the planet and potentially contribute to a resolution to the terminal mass problem of giant planet formation \citep{Morby2014}. A thorough exploration of this possibility is beyond the scope of the semi-analytic calculations presented in this work, and deserves a more detailed treatment, employing 3D magnetohydrodynamic simulations \citep{Gressel2013}.

\subsection{Magnetic Torques}
While the region interior to the truncation radius is envisioned to be devoid of gas, the circumplanetary disk that resides exterior to $R_t$ can be thought of as being largely unperturbed by magnetic stresses, on orbital timescales. In other words, the general picture outlined by the hydrodynamical simulations of \citet{Judit2016} remains valid beyond the planetary magnetosphere. Accordingly, let us now calculate the secular angular momentum exchange between the quasi-Keplerian disk and the planet, facilitated by magnetic induction.

In the $\mathcal{R}_m\gg1$ regime appropriate to the problem at hand, field lines that originate within the planet and vertically puncture the circumplanetary disk, $B_z$, get rapidly\footnote{The associated timescale is of order $\sim2\pi/|\omega-n|$, where $\omega$ is the planetary spin frequency and $n$ is the mean motion.} wound up by Keplerian shear (e.g. \citealt{ArmitageClarke1996}). In other words, the orbital motion of the gas induces an azimuthal magnetic field within the disk, $B_\varphi$, though flux-freezing. Obviously, growth of $B_\varphi$ cannot persist indefinitely and indeed, this process is reset by magnetic reconnection, such that when the pitch angle $\gamma=B_\varphi/B_z$ reaches a value in excess of unity, it is impulsively restored back to zero and the cycle repeats \citep{Uzdensky2002}. 

In light of the short timescale over which the induction-reconnection cycle unfolds, it is sensible to average over it, and adopt a secular value of $\gamma$ that is representative of the mean pitch angle of the field lines within the disk (Figure \ref{pretty}). Following \citet{ArmitageClarke1996,Laietal2011,SpaldingBat2015} and the references therein, we adopt $| \gamma |=1$ everywhere\footnote{We note that there is a radially thin zone on either side of the corotation radius where Keplerian shear is weak enough for Ohmic diffusion to maintain $\gamma\lesssim1$. However, this region is negligibly small, and can be readily ignored \citep{MattPudritz2005}.} in the magnetically connected region of the disk $(r\lesssim0.15\,\Rh)$. Importantly, the sign of $B_\varphi$ is determined by the direction of the differential flow, as measured in a frame co-rotating with the planet. Therefore, disk material residing interior to the corotation radius, $R_c=(\G\,\Mp/\omega^2)^{1/3}$, acts to increase the planetary rotation, while gas exterior to $R_c$ diminishes the planetary spin. 

The magnetic torque exerted upon the planet by the disk is given by the sum of the off-diagonal components of the Maxwell stress tensor \citep{LivioPringle1992}:
\begin{align}
\bigg(\frac{dJ}{dt}\bigg)_m=-\frac{4\pi}{\mu_0}\int B_{\varphi}\,B_z\,r^2\,dr = -\frac{4\pi}{\mu_0}\int\gamma\,\frac{\mathcal{M}^2}{r^4}\,dr,
\label{tau}
\end{align}
where $J$ is the planetary angular momentum, and the integral is envisioned to run over the magnetically connected domain of the disk. Because the integrand in the above equation falls off rapidly with radius, an outer boundary that is sufficiently large for magnetic coupling at distance to become negligible can be treated as being effectively infinite (quantitatively, replacing $0.15\,\Rh$ with $\infty$ as the upper bound of the integral in equation \ref{tau} makes virtually no difference for the system under consideration). Thus, adopting this simplification and accounting for the sign change of the induced field across the corotation radius, we obtain the following expressions:
\begin{align}
\bigg(\frac{dJ}{dt}\bigg)_m &= \begin{dcases} -\frac{4\,\pi}{3\,\mu_0}\frac{\mathcal{M}^2}{R_t^3} & \omega > \sqrt{\frac{\G\,M}{R_t^3}} \\ \frac{4\,\pi}{3\,\mu_0}\frac{\mathcal{M}^2 \, \big(R_c^3-2\,R_t^3 \big)}{R_t^3\,R_c^3} & \omega \leqslant \sqrt{\frac{\G\,M}{R_t^3}}. \end{dcases}
\label{tautrue}
\end{align}

Equation (\ref{tautrue}) has a well defined zero-torque solution, which is simply $R_c=2^{1/3}R_t$ (equivalently, $\omega=\sqrt{\G\,\Mp/2\,R_t^3 }$). The characteristic timescale on which the system approaches this equilibrium state from breakup, $\omega_b$, is:
\begin{align}
\tau_m&=\frac{3\,\mu_0}{4\,\pi}\frac{I\,\Mp\,R_t^3\,\Rp^2\,\omega_b}{\mathcal{M}^2} \approx \frac{2}{I}\Bigg[ \frac{\mu_0\,\sqrt{\G\,\Mp^{15}\,\Rp^7}}{\Mdot^6\,\mathcal{M}^2 } \Bigg]^{1/7},
\label{timescale}
\end{align}
where $I$ is the specific moment of inertia. If we model the interior structure of the young giant planet as a polytropic body with index $\xi=3/2$ (appropriate for a fully convective object), then $I=0.21$ and for our fiducial parameters, $\tau_m\approx2\times10^4\,$years - a timescale two order of magnitude shorter than the typical lifetime of a protoplanetary disk and much smaller than the corresponding value obtained by \citet{TS1996}. We note that the large discrepancy in spin-down timescales obtained here and in previous work stems from the fact that \citet{TS1996} consider a later epoch in Jovian evolutionary history, when magnetic disk-planet coupling is significantly weaker.

Although equation (\ref{timescale}) clearly indicates that the magnetically facilitated planet-disk angular momentum transfer mechanism is exceedingly efficient, we caution that the mere existence of a magnetized circumplanetary disk is insufficient to resolve the terminal spin problem. This is because taken in isolation, the disk's angular momentum budget is much smaller than that of the planet (note that this situation is reversed in the case of circumstellar disks):
\begin{align}
\frac{J_{\rm{disk}}}{J}\lesssim\frac{\int_{0}^{\infty}\int_{R_t}^{\Rh}4\pi\,r\,\rho\,e^\frac{-z^2}{2h^2}\,\sqrt{\G\,\Mp\,r} \, dr\,dz}{I\,\Mp\,\Rp^2\,\omega_b}\sim0.1.
\label{Lratio}
\end{align}
Thus, if the planet and the circumplanetary disk were a closed system, magnetic torques would merely propel the inner regions of the disk to greater orbital radii, while the planetary spin would slow down by a negligible amount. Contrary to this picture, hydrodynamic simulations show that gas is rapidly circulated within the planetary Hill sphere (for example, disk radial velocities of order $\sim 10^{-3}-10^{-2}\,v_{\rm{kep}}$ are observed in the simulations of \citealt{Tanigawa2012,Judit2016}), such that the disk persists in steady-state around the planet, even when magnetic torques are taken into account. As a result, recycling of disk material plays a key role in the successful operation of the magnetic spin-down mechanism, as it connects the circumplanetary disk with its exterior environment and allows the planet to expel its spin angular momentum to the circumstellar nebula.

\subsection{Spin Evolution}

\begin{figure}
\includegraphics[width=\columnwidth]{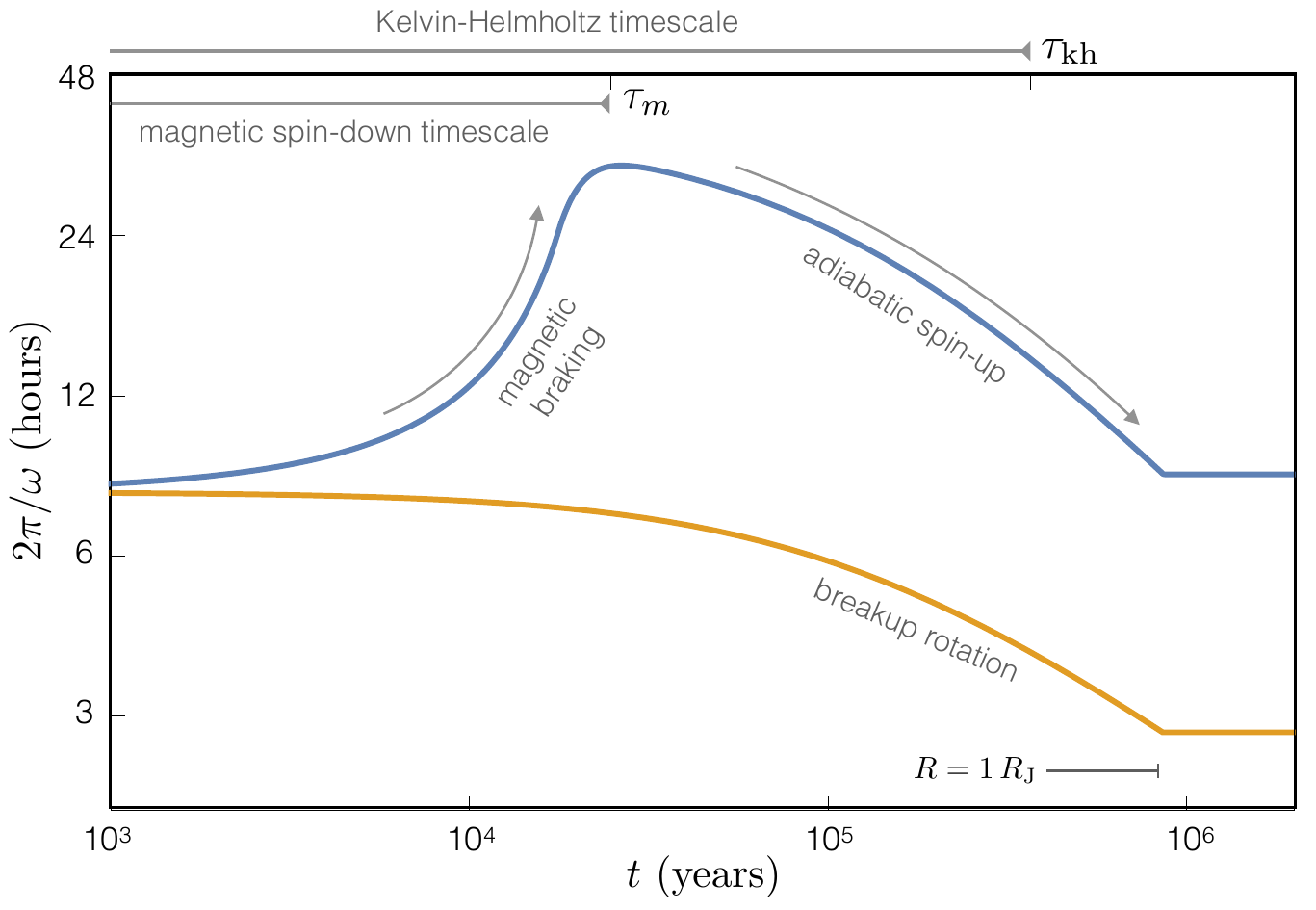}
\caption{Evolution of the the planetary spin, subject to magnetic braking and gravitational contraction. Initialized at breakup velocity, the planet experiences a rapid loss of rotational angular momentum on the timescale, $\tau_m$. Subsequently, as gravitational contraction ensues on the Kelvin-Helmholtz timescale, $\tau_{\rm{kh}}$, the planet spins up adiabatically due to the associated shrinking of the truncation radius, $R_t$. When the physical radius reaches a value equal to that of Jupiter, the spin evolution stops and the planet is left rotating with a period of $2\pi/\omega\sim9\,$hours.}
\label{spinevol}
\end{figure}

In addition to magnetic braking, other effects such as continued accretion and gravitational contraction, may contribute to the rotational evolution of a newly formed giant planet (see e.g. \citealt{BatAd2013}). Drawing upon an analogy with the related case of T-Tauri stars, we note that once the planet becomes luminous enough to develop a strong magnetic field, it is likely that direct accretion of angular momentum can be ignored, since the vertically precipitating flow that can be captured within the apex of the critical field line (i.e. accretion curtain) carries a comparatively small amount of angular momentum\footnote{This assertion is generally validated in the simulation suite of \citet{Judit2016}.}. Gravitational contraction, on the other hand cannot be immediately disregarded as being irrelevant.

The radius evolution of a polytropic body that radiates at a constant effective temperature can be approximately modeled by equating the loss of binding energy $\mathcal{E}=-b\,\G\,\Mp^2/\Rp$ to the surface luminosity:
\begin{align}
\frac{d\mathcal{E}}{dt}=b\frac{\G\,\Mp^2}{\Rp^2}\frac{d\Rp}{dt}=-4\,\pi^2\Rp^2\,\sigma_{\rm{sb}}\,T_{\rm{eff}}^4,
\label{DEdt}
\end{align}
where $b=3/(10-2\xi)=3/7$. Adopting $R_0=2R_{\rm{J}}$ as an initial condition, the solution to the above differential equation is parameterized by the Kelvin-Helmholtz timescale, $\tau_{\rm{kh}}$:
\begin{align}
\Rp&=\bigg(\frac{b\,\G\,\Mp^2\,R_0^3}{b\,\G\,\Mp^2+12\,\pi\,R_0^3\,\sigma_{\rm{SB}}\,T_{\rm{eff}}^4\,t} \bigg)^{1/3}\nonumber \\
&\rightarrow\max\bigg[R_{\rm{J}},R_0\bigg(\frac{\tau_{\rm{kh}}}{\tau_{\rm{kh}}+3\,t} \bigg)^{1/3}\bigg].
\label{Roft}
\end{align}
Importantly, equation (\ref{DEdt}) does not account for degeneracy pressure support, which prevents the planet from contracting below a specific radius, close to that of present-day Jupiter \citep{Stevenson1977}. As a consequence, we impose an artificial limitation upon the above expression, which prevents the radius from diminishing beyond its observed value. We note further that in general, gravitational contraction need not end before dissipation of the circumplanetary disk, implying the possibility of mild post-accretion increase in the rotation rate. 

With all of the necessary ingredients defined, the differential equation for the planetary spin, $\omega$, takes the following form:
\begin{align}
2\,I\,\Mp\,\Rp\,\omega\,\frac{d\Rp}{dt}+I\,\Mp\,\Rp^2\,\frac{d\omega}{dt}=\bigg(\frac{dJ}{dt}\bigg)_m,
\label{omegaoft}
\end{align}
where $\Rp$, $d\Rp/dt$ and $(dJ/dt)_m$ are given by equations (\ref{Roft}), (\ref{DEdt}) and (\ref{tautrue}) respectively. Beyond the assumptions of our rudimentary model, we note that while the characteristic values of $\Mdot$ and $T_{\rm{eff}}$ themselves are somewhat uncertain, their time-evolution is even more obscure since the accretion process can enhance the interior entropy, leading to a broad range of possible luminosity tracks \citep{2016ApJ...819L..14O}. In our numerical solution of this ODE, we circumvent this ambiguity by holding $\Mdot$ and $T_{\rm{eff}}$ constant at their nominal values, while self-consistently evolving the planetary dipole movement as well as the magnetic truncation radius in accord with equations (\ref{Eflux}) and (\ref{Rtr}). Adopting the breakup velocity as a starting condition for $\omega$, Figure (\ref{spinevol}) depicts the derived time-series of the planetary rotation period, over a span of $2\,$Myr. 

Two characteristic phases of evolution are clearly visible in Figure (\ref{spinevol}). First, on a timescale $\tau_m$ (given by equation \ref{timescale}), the system approaches the disk-locked equilibrium state, wherein the planetary spin rate is marginally slower than the orbital frequency at the truncation radius. During subsequent evolution that unfolds on the Kelvin-Helmholtz timescale, the planetary spin adiabatically follows the disk-locked state as gravitational contraction forces the dipole moment to diminish, increasing the rotation rate. Upon reaching the current radius of Jupiter, the physical evolution of the system stops and the rotation period becomes frozen-in at $2\pi/\omega\approx9\,$hours. Note that the dramatic changes in the planetary rotation rate over the simulated timespan are driven entirely by the variations in the magnetospheric truncation radius, highlighting our model's sensitive dependence on $R_t$.

Despite matching the true rotation rate of Jupiter relatively well, we caution that the solution depicted in Figure (\ref{spinevol}) should be viewed as being essentially illustrative rather than exact. Indeed, there exists a number of intricacies and uncertainties that equation (\ref{omegaoft}) simply does not account for. In particular, recall that we have ignored the inevitable decrease of $\Mdot$ that must come about in a waining circumstellar nebula, the timing of giant planet conglomeration relative to disk dissipation, as well as the decline of $T_{\rm{eff}}$ that ensues as the planet contracts. Nevertheless, the presented calculation demonstrates the efficiency of the magnetic braking mechanism within the context of planet-disk interactions, and offers qualitative insight into the physical processes that operate concurrently with giant planet formation and regulate angular momentum exchange within the planet's local sphere of influence. 

\section{Conclusion}\label{sect4}

The question of why despite undergoing rapid accumulation of nebular gas, long-period Jupiter-class object have rotation rates well below breakup is fundamentally important to understanding the concluding phases of giant planet formation \citep{TS1996}. In this work, we have revisited this problem from semi-analytic grounds, and identified an efficient spin-down mechanism that facilitates a time-irreversible transfer of spin angular momentum to the circumstellar nebula via magnetic coupling between the planetary interior and the circumplanetary disk. Crucially, the derived process arises as an inescapable consequence of the planet's enhanced luminosity as well as nebular gap-opening, both of which naturally occur during the runaway accretion phase of conglomeration.

Qualitatively, the following picture is envisioned. When the planetary mass becomes large enough for the Hill radius to exceed the disk scale-height, the planet gravitationally clears out the co-orbital material \citep{Crida2006,Fung2014}, allowing a circumplanetary disk to develop within its sphere of influence \citep{Tanigawa2012,Judit2016}. The inner regions of this disk are characterized by temperatures high enough to render the gas mildly conductive via thermal ionization of Akali metals. Simultaneously, vigorous convection within the newly-formed planet's interior generates a strong magnetic field that truncates the disk, and couples the planetary interior to the quasi-Keplerian flow \citep{Christensen2009}. The circumplanetary disk then rapidly extracts spin angular momentum from the planet, while meridional circulation of gas within the Hill sphere recycles the circumplanetary gas into the circumstellar nebula \citep{ArmitageClarke1996,Uzdensky2002}. Consequently, the observed slow rotation of giant planets is reproduced. Critically, within the framework of the outlined magnetic braking mechanism, the terminal spin is essentially determined by the location of the magnetospheric truncation radius of the circumplanetary disk at the time of dispersal, as well as the physical radius of the planet.

Although the calculations presented in this paper adopt characteristic numbers relevant to a young Jupiter-mass planet, it is worth noting that our rudimentary model does not exhibit an excessively strong dependence upon poorly constrained parameters. In particular, due to the approximately linear and inverse dependency of $\tau_m$ (equation \ref{timescale}) on $M$ and $\Mdot$ respectively, magnetic spin-down time is roughly proportional to the planetary accretion timescale. Meanwhile, the dependence of $\tau_m$ upon the assumed surface field strength is even weaker. Correspondingly, if we reduce $M$ and $\Mdot$ by a factor of $\sim3$ while adopting a somewhat smaller value of $T_{\rm{eff}}=1000\,$K and take $\sim0.8R_{\rm{J}}$ as the lower bound in equation (\ref{Roft}) in accord with nominal parameters of Saturn, we obtain a very similar spin-evolution to that depicted in Figure (\ref{spinevol}), characterized by a terminal rotation period of $\sim11\,$hours.

While the approximate model developed in this work successfully captures the key physical processes at play, its semi-analytical nature warrants further examination of the problem. In particular, the intriguing possibility that the generation of a strong magnetic field by a highly luminous planet can substantially reduce its accretion cross-section by diverting in-falling material through magnetic stresses \citep{AG2012}, deserves to be evaluated with the aid of detailed numerical simulations. Correspondingly, if the development of a $B_{\rm{s}}\sim0.1-1$ kiloGauss field can indeed act to quench the planetary growth (which in turn determines the luminosity), our envisioned angular momentum extraction mechanism would be rendered self-limiting.

In addition to further theoretical implications, the precise parameter range and evolutionary timespan over which the envisioned scenario applies, remain to be determined. Crudely, our simple model suggests that planets with effective temperatures considerably cooler than $T_{\rm{eff}}\lesssim1000\,$K, would have disks that are insufficiently ionized for significant magnetic coupling to ensue. Meanwhile, radiative hydrodynamics simulations of \citet{Judit2017} suggest that circumplanetary disks that encircle planets more massive than $\Mp\gtrsim5\, M_{\rm{J}}$ can develop significant eccentricities, further complicating the picture. Thus, the calculations outlined within this work provide an important stepping stone towards the full resolution of the terminal spin problem of core-nucleated accretion theory, and motivate continued exploration of magnetohydrodynamic effects within the context of the final stages of giant planet formation. 

\acknowledgments
\textbf{Acknowledgments}. I am thankful to Fred Adams, Greg Laughlin, Mike Brown and Alessandro Morbidelli for illuminating discussions, as well as to Dan Tamayo for providing a thorough and insightful referee report. Figure 2 was drawn by James Tuttle Keane under contract from California Institute of Technology. I am further grateful to the David and Lucile Packard Foundation for their generous support.

\end{document}